\documentclass[a4paper]{article}

\usepackage{INTERSPEECH2020}
\usepackage{hyperref}
\usepackage[detect-all]{siunitx}
\usepackage{mathtools}
\usepackage{comment}
\usepackage{multirow}
\usepackage[subrefformat=parens]{subcaption}

\usepackage{xspace}
\makeatletter
\DeclareRobustCommand\onedot{\futurelet\@let@token\@onedot}
\def\@onedot{\ifx\@let@token.\else.\null\fi\xspace}

\def\eg{\emph{e.g}\onedot} 
\def\ie{\emph{i.e}\onedot}

\def\etal{\emph{et al}\onedot}
\makeatother

\newcommand{\trans}[1]{#1^\mathsf{T}}

\DeclareMathOperator*{\argmin}{arg\,min}\def\appendixautorefname~#1\null{~#1 \null}

\newcommand{\tablescale}{0.85}

\newcommand{\myparagraph}[1]{\vspace{1mm}\noindent\textbf{#1}:}

\title{End-to-End Speaker Diarization for an Unknown Number of Speakers\\with Encoder-Decoder Based Attractors}
\name{Shota Horiguchi$^1$, Yusuke Fujita$^1$, Shinji Watanabe$^2$, Yawen Xue$^1$, Kenji Nagamatsu$^1$}
\address{
  $^1$Hitachi, Ltd.\qquad
  $^2$Johns Hopkins University}

\email{\{shota.horiguchi.wk, yusuke.fujita.su, yawen.xue.wn\}@hitachi.com, shinjiw@ieee.org}

\begin{document}
\abovedisplayskip=4pt
\belowdisplayskip=4pt
\setlength\floatsep{11pt}
\setlength\textfloatsep{11pt}
\setlength\abovecaptionskip{5pt}
\setlength\dblfloatsep{5pt}
\setlength\dbltextfloatsep{11pt}

\maketitle
\begin{abstract}
  End-to-end speaker diarization for an unknown number of speakers is addressed in this paper.
  Recently proposed end-to-end speaker diarization outperformed conventional clustering-based speaker diarization, but it has one drawback: it is less flexible in terms of the number of speakers.
  This paper proposes a method for encoder-decoder based attractor calculation (EDA), which first generates a flexible number of attractors from a speech embedding sequence.
  Then, the generated multiple attractors are multiplied by the speech embedding sequence to produce the same number of speaker activities.
  The speech embedding sequence is extracted using the conventional self-attentive end-to-end neural speaker diarization (SA-EEND) network.
  In a two-speaker condition, our method achieved a \SI{2.69}{\percent} diarization error rate (DER) on simulated mixtures and a \SI{8.07}{\percent} DER on the two-speaker subset of CALLHOME, while vanilla SA-EEND attained \SI{4.56}{\percent} and \SI{9.54}{\percent}, respectively. In unknown numbers of speakers conditions, our method attained a \SI{15.29}{\percent} DER on CALLHOME, while the x-vector-based clustering method achieved a \SI{19.43}{\percent} DER.
\end{abstract}

\section{Introduction}
Speaker diarization is the task to estimate ``who spoke when'' from an audio recording.
It is a key technology for various applications using automatic speech recognition (ASR) in multi-talker scenarios such as telephone conversations \cite{kenny2010diarization}, meetings \cite{anguera2007acoustic}, conferences and lectures \cite{zhu2007multi}, TV shows \cite{vallet2012multimodal}, and movies \cite{kapsouras2017multimodal}.
Accurate diarization has been proven to improve ASR performance by constraining a speech mask when constructing a beamformer for speech separation \cite{kanda2019guided,zorila2019investigation}.

One major approach for speaker diarization is the clustering-based method \cite{shum2013unsupervised,sell2014speaker}, which applies the following processes to an input audio one by one: speech activity detection, speech segmentation, feature extraction, and clustering.
Progress on better speaker embeddings, such as x-vectors \cite{snyder2019speaker,diez2019bayesian} and d-vectors \cite{wang2018speaker,zhang2019fully}, have enabled accurate clustering-based diarization.
However, most clustering-based approaches (except for a few studies, \eg, \cite{huang2020speaker}) cannot deal with speaker overlap because each time slot is assigned to one speaker.

End-to-end speaker diarization called EEND \cite{fujita2019end1,fujita2019end2} has been proposed to overcome this situation.
The EEND is optimized to calculate diarization results for every speaker in a mixture from input audio features using permutation invariant training (PIT) \cite{yu2017permutation}.
The EEND, especially self-attentive EEND (SA-EEND), showed the effectiveness of end-to-end training of the diarization model by outperforming conventional clustering-based methods.
One drawback it has is that the maximum number of speakers is pre-determined by the network architecture, and it cannot deal with a case where the number of speakers is higher.
On this point, EEND is less flexible than clustering-based methods, where the number of speakers can be easily changed by setting the number of clusters during inferences.

This paper proposes an encoder-decoder based attractor calculation method called EDA.
It determines a flexible number of---and theoretically an infinite number of attractors---from a speech embedding sequence.
We applied it to SA-EEND to enable diarization with a flexible number of speakers.
Then, the diarization results are calculated using dot products between all pairs of attractors and embeddings.
Evaluation results on both simulated mixtures and real recordings showed that our method achieved better results with both fixed and unknown numbers of speakers than the x-vector-based clustering method and conventional SA-EEND.

\section{Related work}
Several methods in the context of speech separation can process speech mixtures of a flexible number of speakers.
One series of methods involve applying the one-vs-rest approach iteratively \cite{kinoshita2018listening,shi2018listen,von2019all,takahashi2019recursive}.
However, it has a major drawback in that the calculation is conducted until all the speakers are extracted, so the computational time increases linearly as the number of speakers increases.
Another series involve attractor-based approaches including Deep Attractor Network (DANet) \cite{chen2017deep}. 
It does not limit the number of speakers in the inference phase; however, the number of speakers has to be known a priori.
Anchored DANet \cite{luo2018speaker} successfully solved the aforementioned problems, but it always requires calculating dot products between all the possible selections of anchors and extracted embeddings even in the inference phase.
Thus, it is not scalable in terms of the number of speakers.

Several efforts have been made to calculate representatives from an embedding sequence in an end-to-end manner.
Lee \etal proposed Set Transformer to implement set-to-set transformation \cite{lee2019set}, but the number of outputs has to be defined beforehand.
Meier \etal implemented end-to-end clustering by estimating the distribution for every possible number of clusters $K\in\{1,\dots,K_\mathrm{max}\}$ \cite{meier2018learning} so that the maximum number is limited by the network architecture.
Li \etal proposed encoder-decoder based clustering for speaker diarization \cite{li2019discriminative}, which is the most related to EDA. 
However, the output is a sequence of cluster numbers of each input, so each time slot is assigned to one cluster; therefore, it cannot deal with speaker overlap.
Our proposed EDA, in contrast, determines a flexible number of attractors from an embedding sequence without prior knowledge about the number of clusters.

\section{End-to-end neural diarization: Review}
Here we briefly introduce our end-to-end diarization framework named EEND \cite{fujita2019end1,fujita2019end2}.
The EEND takes a $T$-length sequence of log-scaled Mel-filterbank based features as an input, and processes it using bi-directional long short-term memory (BLSTM) \cite{fujita2019end1} or Transformer encoders \cite{fujita2019end2} to obtain an embedding $\mathbf{e}_t\in\mathbb{R}^D$ at each time slot.
After that, a linear transformation $f:\mathbb{R}^D\to\mathbb{R}^S$ with an element-wise sigmoid function is applied to calculate posteriors $\hat{\mathbf{y}}_t=[\trans{\hat{y}_{t,1},\dots,\hat{y}_{t,S}]}\in(0,1)^S$ of $S$ speakers at time slot $t$.
In the training phase, the EEND is optimized using the PIT scheme \cite{yu2017permutation}, \ie, the loss is calculated between $\hat{\mathbf{y}}_t$ and the groundtruth labels $\mathbf{y}_t=[\trans{y_{t,1},\dots,y_{t,S}]}\in\{0,1\}^S$ as follows:
\begin{align}
    L_d&=\frac{1}{TS}\argmin_{\phi\in\mathrm{perm}(1,\dots,S)}\sum_{t=1}^{T}H\left(\mathbf{y}_t^\phi,\hat{\mathbf{y}}_t\right),
    \label{eq:diarization_loss}
\end{align}
where $\mathrm{perm}(1,\dots,S)$ is the set of all the possible permutation of speakers, $\mathbf{y}_t^\phi\in\{0,1\}^S$ is the permuted labels at $t$, and $H(\mathbf{y}_t,\hat{\mathbf{y}}_t)$ is the binary cross entropy determined as follows:
\begin{align}
    H\left(\mathbf{y}_t,\hat{\mathbf{y}}_t\right)&\coloneqq\sum_s{-y_{t,s}\log{\hat{y}_{t,s}}-\left(1-y_{t,s}\right)\log{\left(1-\hat{y}_{t,s}\right)}}.
    \label{eq:cross_entropy}
\end{align}

\section{Proposed method}
\begin{figure}[t]
    \centering
    \includegraphics[width=\linewidth]{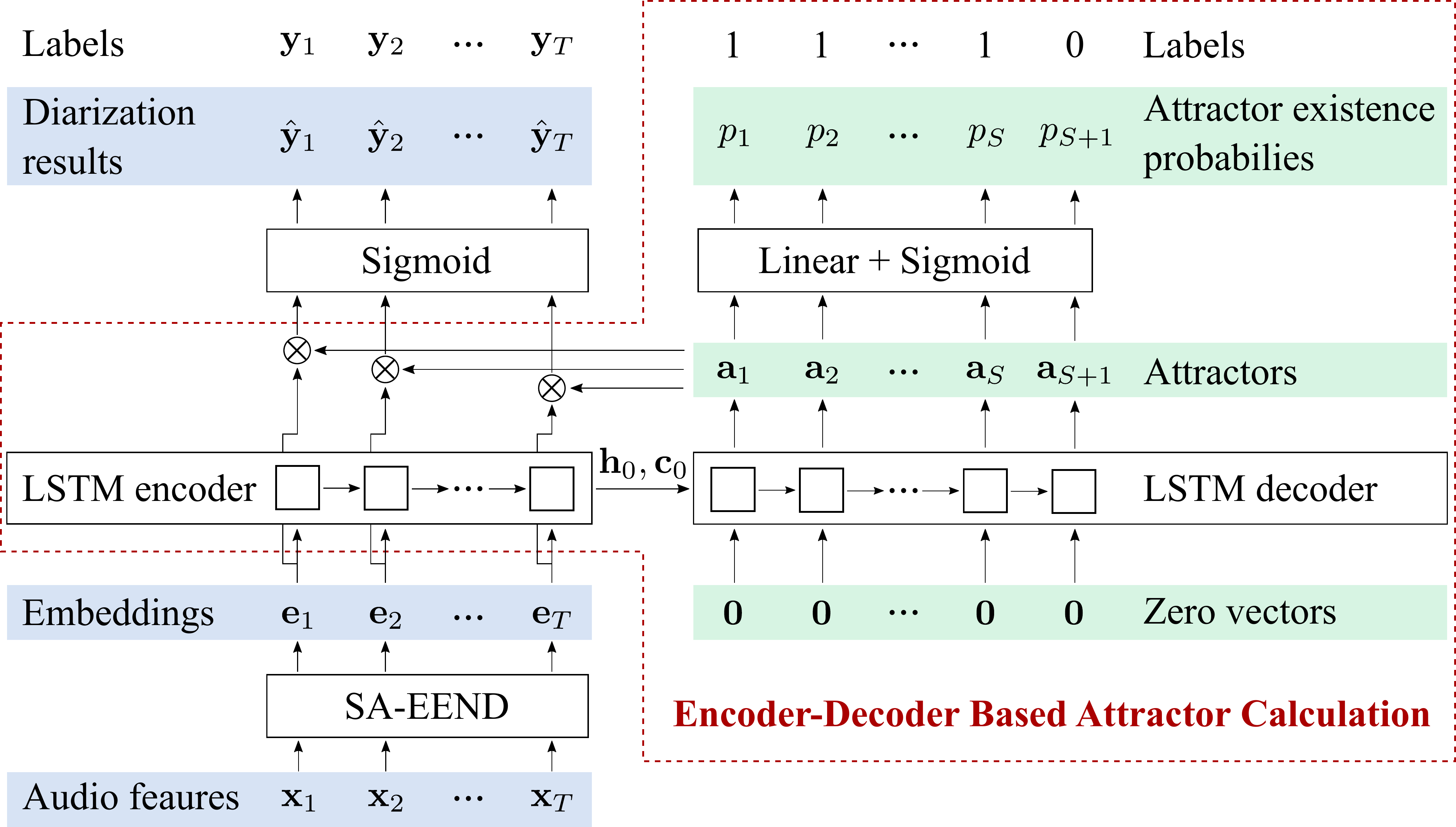}
    \caption{SA-EEND with encoder-decoder based attractor calculation.}
    \label{fig:EDA_diarization}
\end{figure}

The EEND has a critical problem, in that the output size is limited by the network architecture; the linear transformation $f$ restricts the number of speakers $S$ during inference. Therefore, it cannot deal with a case where the input mixture contains a higher number of speakers than the capacity.
Therefore, we utilized an attractor-based method.
To make our method end-to-end trainable, we designed Encoder-Decoder based Attractor calculation (EDA) to determine attractors from an embedding sequence.
The overview of our proposed method is shown in \autoref{fig:EDA_diarization}.
We used the same self-attentive network in \cite{fujita2019end2} as a backbone to obtain an embedding $\mathbf{e}_t$ at each time slot.
In this section, we explain how we calculate a flexible number of attractors from the embeddings and obtain diarization results using the attractors.

\subsection{Encoder-decoder based attractor calculation}
\label{sec:eda}

To calculate a flexible number of attractor points from variable lengths of embedding sequences, we utilize LSTM-based encoder-decoder \cite{sutskever2014sequence}.
A sequence of $D$-dimensional embeddings $(\mathbf{e}_t)_{t=1}^{T}$ is fed into the unidirectional LSTM encoder, obtaining the final hidden state embedding $\mathbf{h}_0\in\mathbb{R}^D$ and the cell state $\mathbf{c}_0\in\mathbb{R}^D$:
\begin{align}
	\mathbf{h}_0,\mathbf{c}_0=\mathrm{Encoder}\left(\mathbf{e}_1,\cdots,\mathbf{e}_T\right).
\end{align}
Next, time-invariant $D$-dimensional attractors $(\mathbf{a_s})_s$ are calculated using an unidirectional LSTM decoder with the initial states $\mathbf{h}_0$ and $\mathbf{c}_0$ as follows.
\begin{align}
    \mathbf{h}_s,\mathbf{c}_s,\mathbf{a}_s=\mathrm{Decoder}\left(\mathbf{h}_{s-1}, \mathbf{c}_{s-1},\mathbf{0}\right)
\end{align}
We use a $D$-dimensional zero vector $\mathbf{0}$ as the input for the decoder at each decoding step.
Theoretically infinite numbers of attractors can be calculated using the LSTM decoder.
The probability of whether or not the attractor $\mathbf{a}_s$ exists to determine when to stop the attractor calculation is computed using a fully-connected layer with a sigmoid function as
\begin{align}
	p_s=\frac{1}{1+\exp{\left(-\left(\trans{\mathbf{w}}\mathbf{a}_s+b\right)\right)}},
\end{align}
where $\mathbf{w}$ and $b$ are the trainable weights and bias of the fully-connected layer, respectively.

We note that the output attractors $(\mathbf{a}_s)_s$ depend on the order of the input embeddings $(\mathbf{e}_t)_{t=1}^{T}$ because we use LSTMs for the EDA.
To investigate the effect of the input order, we used two types of embedding order.
One was a chronological order, \ie, the embeddings were sorted by time slot indexes.
The other was a shuffled order. In this case, we used a shuffled order of embeddings, namely $(\mathbf{e}_{\psi(t)})_{t=1}^{T}$, where $(\psi(1),\dots,\psi(T))$ is one of the permutations of $(1,\dots,T)$, for the input to the EDA.

In the training phase, we defined the groundtruth labels $\mathbf{l}=\trans{[l_1,\dots,l_{S+1}]}$ using the actual number of speakers $S$ as follows:
\begin{align}
	l_s=\begin{cases}
	1 & \left(s\in\left\{1,\dots,S\right\}\right)\\
	0 & \left(s=S+1\right).
	\end{cases}
\end{align}
Also the attractor existence loss $L_a$ between the labels and the estimated probabilities $\mathbf{p}=\trans{[p_1,\dots,p_{S+1}]}$ were calculated using the binary cross entropy in \autoref{eq:cross_entropy} as
\begin{align}
    L_a&=\frac{1}{1+S}H\left(\mathbf{l},\mathbf{p}\right).
    \label{eq:attractor_existence_loss}
\end{align}

In the inference phase, if the number of speakers $S$ was given, we use the first $S$ attractors, which were the output from the EDA.
If the number of speakers was unknown, we first estimated it using
\begin{align}
    \hat{S}=\max\left\{s\mid s\in\mathbb{Z}_{+}\land p_s\geq\tau\right\}
    \label{eq:n_attractor_estimation}
\end{align}
with a given threshold $\tau$ and then used the first $\hat{S}$ attractors.

\subsection{Speaker diarization using EDA}
We respectively define the matrix formulations of the embeddings extracted from the SA-EEND and the attractors from the EDA as follows.
\begin{align}
    E&\coloneqq\left[\mathbf{e}_1,\dots,\mathbf{e}_T\right]\in\mathbb{R}^{D\times T}\\
    A&\coloneqq\left[\mathbf{a}_1,\dots,\mathbf{a}_S\right]\in\mathbb{R}^{D\times S}
\end{align}
The posterior probabilities can be calculated using the inner product of every embedding-attractor pair as follows:
\begin{align}
    \hat{Y}=\sigma(\trans{A}E)\in\left(0,1\right)^{S\times T},
\end{align}
where $\sigma(\cdot)$ is the element-wise sigmoid function.
Note that the output size was determined using the number of attractors so that our method could output the diarization results of a flexible number of speakers.
Finally, diarization loss was calculated in the same way as SA-EEND using the PIT found in \autoref{eq:diarization_loss}.

The total loss is defined by the diarization loss in \autoref{eq:diarization_loss} and the attractor existence loss in \autoref{eq:attractor_existence_loss} as follows:
\begin{align}
    L=L_d+\alpha L_a,
\end{align}
where $\alpha$ is the weighting parameter.
In this study, $\alpha$ was set to $1.0$ when the simulated data were used for training and $0.01$ for adaptation on real datasets.

\section{Experiments}
\subsection{Data}
For the training and evaluation, we used simulated mixtures created from Switchboard-2 (Phase I \& I\hspace{-.1em}I \& I\hspace{-.1em}I\hspace{-.1em}I), Switchboard Cellular (Part 1 \& 2), and the NIST Speaker Recognition Evaluation (2004 \& 2005 \& 2006 \& 2008) for speech and the MUSAN corpus \cite{snyder2015musan} for noise with simulated room impulse responses used in \cite{ko2017study} following the procedure in \cite{fujita2019end2}.
We note that the speaker sets for the training and test datasets were not overlapped.
In \cite{fujita2019end2}, only the 2-speaker dataset was constructed.
In this study, we created 1-, 3-, and 4-speaker datasets with similar overlap ratios $\rho$ to the 2-speaker mixtures.
We also used the telephone conversation dataset CALLHOME (CH) \cite{callhome}, the dialogue recordings from the Corpus of Spontaneous Japanese (CSJ) \cite{maekawa2003corpus}, and the dataset used for the second DIHARD challenge \cite{ryant2019second} to evaluate the performance on real recordings.
The statistics of the datasets used are summarized in \autoref{tbl:dataset}.

\begin{table}[t]
    \centering
    \caption{Dataset to train and test our diarization models.}
    \label{tbl:dataset}
    \subcaption{Simulated datasets.}
    \scalebox{\tablescale}{
    \begin{tabular}{@{}lccc@{}}
        \toprule
        Dataset & \#Spk & \#Mixtures & Overlap ratio $\rho$ (\%)\\\midrule
        \textbf{Train}\\
        \hspace{5mm}Sim1spk& 1 & 100,000&0.0\\
        \hspace{5mm}Sim2spk& 2 & 100,000&34.1\\
        \hspace{5mm}Sim3spk& 3 & 100,000&34.2\\
        \hspace{5mm}Sim4spk& 4 & 100,000&31.5\\\midrule
        \textbf{Test}\\
        \hspace{5mm}Sim1spk & 1 & 500 & 0.0\\
        \hspace{5mm}Sim2spk& 2 & 500/500/500&34.4/27.3/19.6\\
        \hspace{5mm}Sim3spk& 3 & 500/500/500& 34.7/27.4/19.1\\
        \hspace{5mm}Sim4spk& 4 & 500&32.0\\
        \bottomrule
    \end{tabular}
    }\\
    \vspace{8pt}
    \subcaption{Real datasets.}
    \scalebox{\tablescale}{
    \begin{tabular}{@{}lccc@{}}
        \toprule
        Dataset & \#Spk & \#Mixtures & Overlap ratio $\rho$ (\%)\\\midrule
        \textbf{Train}\\
        \hspace{5mm}CALLHOME \cite{callhome}& 2 & 155 & 14.0\\
        \hspace{5mm}CALLHOME \cite{callhome}&3&61&19.6\\
        \hspace{5mm}CALLHOME \cite{callhome}&2-7&249&17.0\\
        \hspace{5mm}DIHARD dev \cite{ryant2019second}&1-10&192&9.8\\\midrule
        \textbf{Test}\\
        \hspace{5mm}CALLHOME \cite{callhome} &2 & 148 & 13.1\\
        \hspace{5mm}CALLHOME \cite{callhome} &3 & 74 & 17.0\\
        \hspace{5mm}CALLHOME \cite{callhome} &2-6 & 250 & 16.7\\
        \hspace{5mm}CSJ \cite{maekawa2003corpus} &2 & 54 & 20.1\\
        \hspace{5mm}DIHARD eval \cite{ryant2019second} &1-9 &194&8.9\\
        \bottomrule
    \end{tabular}
    }
\end{table}

\subsection{Experimental settings}
We basically followed the training protocol of the best model described in \cite{fujita2020endtoend}\footnote{SA-EEND is available at \url{https://github.com/hitachi-speech/EEND}. We will release the source code of SA-EEND with EDA at the same repository.}.
We used SA-EEND with four-stacked Transformer encoders as a baseline and a backbone of our method.
The inputs for the SA-EEND were 345-dimensional log-scaled Mel-filterbank based features, which were also the same as those used in the original paper.
For our method, we extracted a sequence of 256-dimensional embeddings after the last layer normalization \cite{ba2016layer} of the SA-EEND, and fed them into the EDA to calculate attractors.
The threshold $\tau$ in \autoref{eq:n_attractor_estimation} to determine whether or not the attractor existed was set to 0.5.
As we explained in \autoref{sec:eda}, we used two types of input order for the EDA: chronological order and shuffled order.
Unless otherwise noted, we used the same type of order in the training and inference phases.

In this paper, we evaluated our method under the following two conditions: a fixed number of speakers and a flexible number of speakers.
For the fixed number of speakers, we first trained our model using Sim2spk with $\rho=\SI{34.1}{\percent}$ or Sim3spk with $\rho=\SI{34.2}{\percent}$ for 100 epochs.
We used the Adam optimizer \cite{kingma2015adam} with the learning rate schedule proposed in \cite{vaswani2017attention} with warm-up steps of 100,000.
We also finetuned those models using subsets of corresponding numbers of speakers from CALLHOME data to evaluate the performance on the real recordings.
For comparison, the performance on i-vectors or x-vectors using agglomerative hierarchical clustering with probabilistic linear discriminate analysis (PLDA) scoring according to Kaldi's pretrained model \cite{snyder2018xvectors} was also evaluated. In these cases, TDNN-based speech activity detection \cite{peddinti2015jhu} and the oracle number of speakers were used for the evaluation.
For experiments on the flexible speaker condition, we finetuned the 2-speaker model trained on Sim2spk on the concatenation of Sim1spk, Sim2spk, Sim3spk, and Sim4spk for 25 epochs.
We finetuned the model using CALLHOME or DIHARD dev to evaluate the performance on real datasets.
The x-vector-based methods based on the oracle number of speakers and the clustering threshold determined using the training set were also evaluated.

For the evaluation metric, we used the diarization error rate (DER).
The \SI{0.25}{\s} of collar tolerance was defined at the start and end of each segment for the evaluation on the simulated datasets and the CALLHOME dataset.
For the DIHARD dataset, we also used the Jaccard error rate (JER), and we did not use collar tolerance, following the regulation of the second DIHARD challenge \cite{ryant2019second}.

\subsection{Results on a fixed number of speakers}
\begin{table}[t]
    \centering
    \caption{DERs (\%) on 2-speaker datasets.}
    \label{tbl:results_2spk}
    \resizebox{\linewidth}{!}{
    \begin{tabular}{@{}lccccc@{}}
        \toprule
        &\multicolumn{3}{c}{Sim2spk}&\multicolumn{2}{c}{Real}\\\cmidrule(lr){2-4}\cmidrule(l){5-6}
        Method&$\rho=\SI{34.4}{\percent}$&$\SI{27.3}{\percent}$&$\SI{19.6}{\percent}$&CH&CSJ\\\midrule
        i-vector clustering&33.74&30.93&25.96&12.10&27.99\\
        x-vector clustering&28.77&24.46&19.78&11.53&22.96\\
        BLSTM-EEND \cite{fujita2019end1}& 12.28&14.36&19.69&26.03&39.33\\
        SA-EEND \cite{fujita2019end2} & 4.56&4.50&3.85&9.54&20.48\\
        SA-EEND + EDA (Chronol.) & 3.07&2.74&3.04&8.24&18.89\\
        SA-EEND + EDA (Shuffled) & {\bfseries 2.69} &\textbf{2.44}&\textbf{2.60}&\textbf{8.07}&\textbf{16.27}\\
        \bottomrule
    \end{tabular}
    }
\end{table}
\begin{table}[t]
    \centering
    \caption{DERs (\%) on 3-speaker datasets.}
    \label{tbl:results_3spk}
    \resizebox{\linewidth}{!}{
    \begin{tabular}{@{}lcccc@{}}
        \toprule
        &\multicolumn{3}{c}{Sim3spk}&Real\\\cmidrule(lr){2-4}\cmidrule(l){5-5}
        Method&$\rho=\SI{34.7}{\percent}$&$\SI{27.4}{\percent}$&$\SI{19.1}{\percent}$&CH\\\midrule
        x-vector clustering&31.78&26.06&19.55&19.01\\
        SA-EEND &8.69&7.64&6.92&14.00\\
        SA-EEND + EDA (Chronol.) &13.02&11.65&10.41&15.86\\
        SA-EEND + EDA (Shuffled) &\textbf{8.38}&\textbf{7.06}&\textbf{6.21}&\textbf{13.92}\\
        \bottomrule
    \end{tabular}
    }
\end{table}

\begin{table*}[t]
    \centering
    \caption{DERs on Sim2spk ($\rho=\SI{34.4}{\percent}$) using various types of sequences.}
    \label{tbl:various_sequence}
    \resizebox{\linewidth}{!}{
    \begin{tabular}{@{}lcccccccccccc@{}}
        \toprule
        &\multicolumn{2}{c}{Use whole sequence}&\multicolumn{5}{c}{Subsample $1/N$}&\multicolumn{5}{c}{Use the last $1/N$}\\\cmidrule(lr){2-3}\cmidrule(lr){4-8}\cmidrule(l){9-13}
        Method&Chronol. & Shuffled & $N=2$ & $N=4$ & $N=8$ & $N=16$ & $N=32$ & $N=2$ & $N=4$ & $N=8$& $N=16$ & $N=32$\\\midrule
        SA-EEND + EDA (Chronol.)&3.07&30.04&3.54&7.32&14.48&21.13&27.18&3.67&4.97&5.40&6.11&7.68\\
        SA-EEND + EDA (Shuffled)&2.69&2.69&2.70&2.68&2.79&3.09&5.08&3.36&5.92&7.46&8.59&10.65\\
        \bottomrule
    \end{tabular}
    }
\end{table*}

First, we evaluated our method on the 2-speaker condition like the one in \cite{fujita2019end1,fujita2019end2}.
The results are shown in \autoref{tbl:results_2spk}.
The best DERs were attained using EDA trained on shuffled embeddings.
When the model was trained using embeddings in chronological order, the DERs slightly degraded. 
We also show the results on the 3-speaker condition in \autoref{tbl:results_3spk}.
Our method with shuffled embeddings achieved better DERs compared with the conventional x-vector clustering and vanilla SA-EEND.

\myparagraph{Effect of the input order}
To better understand the EDA, we evaluated the diarization performance on both chronologically-ordered sequences and shuffled sequences. We also tried to reduce the length of sequences by subsampling embeddings or using the last $1/N$ of the sequences.
The results on Sim2spk ($\rho=\SI{34.4}{\percent}$) are shown in \autoref{tbl:results_3spk}. 
When the EDA was trained on chronologically-ordered embeddings, it worked better on chronologically-ordered embeddings but degraded shuffled embeddings.
If the embeddings were subsampled, the performance degradation was also severe even if the samples were ordered chronologically, while using the last $1/N$ could suppress the performance degradation.
These results were that the model captured speech length tendency to output attractors.
However, when the EDA was trained on shuffled embeddings, the model was not affected very much by the order and subsampling. 
These results show that the EDA could capture the overall sequence successfully. 

\myparagraph{Visualization}
\begin{figure}
    \centering
    \begin{minipage}[c]{.49\linewidth}
        \includegraphics[width=\linewidth]{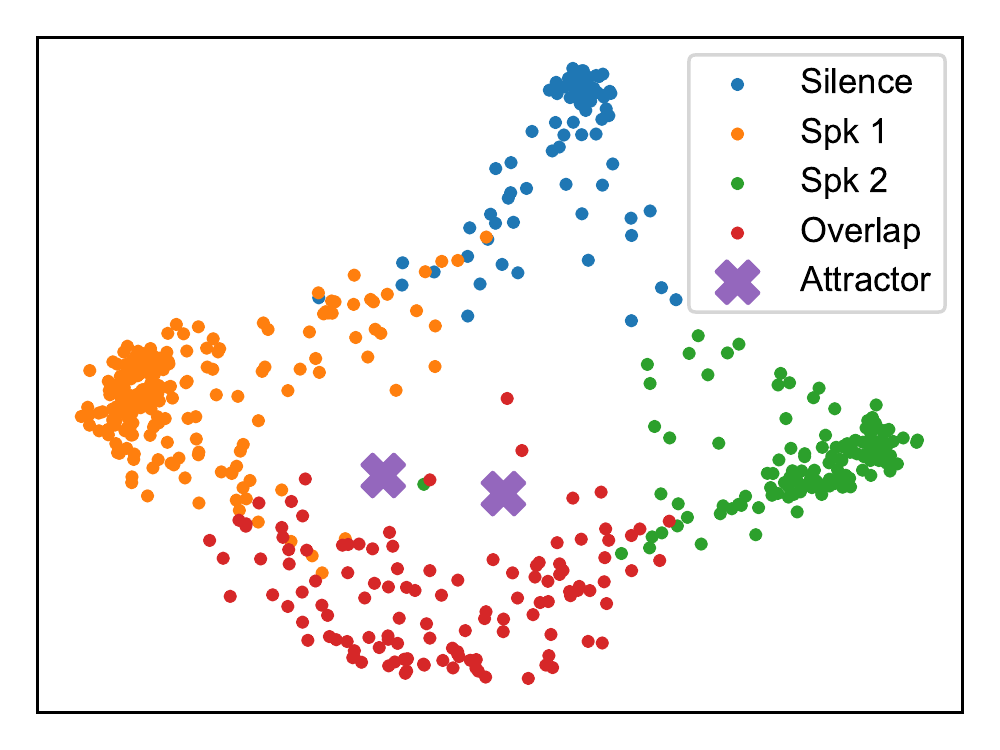}
    \end{minipage}
    \hfill
    \begin{minipage}[c]{.49\linewidth}
        \includegraphics[width=\linewidth]{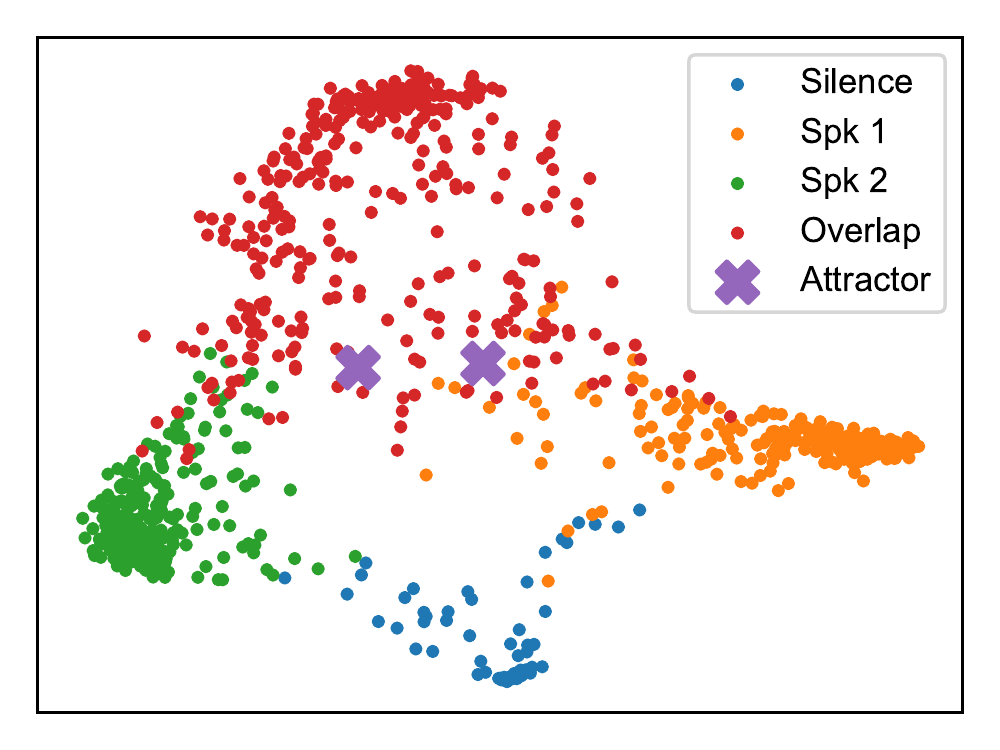}
    \end{minipage}
    \caption{Visualization of embeddings and attractors on 2-speaker mixtures in Sim2spk ($\rho=\SI{34.4}{\percent}$).}
    \label{fig:visualization}
\end{figure}
In \autoref{fig:visualization}, we visualized embeddings and attractors of 2-speaker mixtures by applying PCA to reduce their dimensionality. The embeddings of two speakers were well separated from the silent region, and those of overlapping regions were distributed between two clusters. Attractors were successfully calculated for each of the two speakers.

\subsection{Results on a flexible number of speakers}
We also evaluated our method on a condition involving a flexible number of speakers.
In this case, the order of the embeddings was always shuffled. 
The model was first finetuned from the weights trained on Sim2spk and evaluated on simulated mixtures of a flexible number of speakers.
The results are shown in \autoref{tbl:flex_simulated}.
Our method achieved better DERs than the x-vector clustering-based method.
It achieved \SI{4.33}{\percent} and \SI{8.94}{\percent} DERs on two- and three-speaker mixtures, which showed only 1.64 and 0.56 point degradation from two- or three-speaker specific models, respectively.
Furthermore, our method further improved performance when the actual number of speakers was given, while x-vector clustering worsened performance in most cases using the oracle number of speakers.

\begin{table}[t]
    \centering
    \caption{DERs (\%) on simulated mixtures of a flexible number of speakers.}
    \label{tbl:flex_simulated}
    \scalebox{\tablescale}{
    \begin{tabular}{@{}lcccc@{}}
        \toprule
        &Sim1spk&Sim2spk&Sim3spk&Sim4spk\\\cmidrule(lr){2-2}\cmidrule(lr){3-3}\cmidrule(lr){4-4}\cmidrule(l){5-5}
        Method &  $\rho=\SI{0.0}{\percent}$ & $\SI{34.4}{\percent}$ & $\SI{34.7}{\percent}$ & $\SI{32.0}{\percent}$\\\midrule
        x-vector clustering\\
        \hspace{5mm}Threshold&37.42&7.74&11.46&22.45\\
        \hspace{5mm}Oracle \#Spk&1.67&28.77&31.78&35.76\\
        SA-EEND + EDA\\
        \hspace{5mm}Estimated \#Spk&0.39&4.33&8.94&13.76\\
        \hspace{5mm}Oracle \#Spk&0.16&4.26&8.63&13.31\\
        \bottomrule
    \end{tabular}
    }
\end{table}

We also evaluated our method with real conversations using the CALLHOME.
In this case, the model was finetuned again using the CALLHOME training set and evaluated on the test set.
The results are shown in \autoref{tbl:flex_real}.
Our method achieved a \SI{15.29}{\percent} DER, which outperformed the clustering-based method.
However, it did not perform well when the number of speakers was higher than four.
This is because the CALLHOME contains only ten recordings that include more than four speakers.

\begin{table}[t]
    \centering
    \caption{DERs (\%) on CALLHOME of a flexible number of speakers.}
    \label{tbl:flex_real}
    \resizebox{\linewidth}{!}{
    \begin{tabular}{@{}lcccccc@{}}
        \toprule
        &\multicolumn{6}{c}{\#Spk}\\\cmidrule(l){2-7}
        Method &  2 & 3 & 4 & 5 & 6 & All\\\midrule
        x-vector clustering\\
        \hspace{5mm}Threshold&15.45&18.01&22.68&31.40&34.27&19.43\\
        \hspace{5mm}Oracle \#Spk&8.93&19.01&24.48&32.14&34.95&18.98\\
        SA-EEND + EDA\\
        \hspace{5mm}Estimated \#Spk & 8.50 & 13.24 & 21.46 & 33.16 & 40.29 & \textbf{15.29}\\
        \hspace{5mm}Oracle \#Spk& 8.35 & 13.20 & 21.71 & 33.00 & 41.07 & 15.43\\
        \bottomrule
    \end{tabular}
    }
\end{table}

\begin{table}[tb]
    \centering
    \caption{DERs and JERs (\%) on DIHARD eval.}
    \label{tbl:dihard}
    \scalebox{\tablescale}{
    \begin{tabular}{@{}lcc@{}}
        \toprule
        Method&DER&JER\\\midrule
        DIHARD-2 baseline \cite{sell2018diarization}& 40.86&66.60\\
        Best pre-is2019-deadline \cite{novoselov2019speaker}&35.10&57.11\\
        Best post-is2019-deadline \cite{landini2020but}& 27.11&49.07\\
        SA-EEND + EDA (Estimated \#Speakers)& 32.59&55.99\\
        \bottomrule
    \end{tabular}
    }
\end{table}

Finally, we evaluated our method on the DIHARD dataset.
The evaluation follows the DIHARD 2019 track 2, where speech activity detection has to be conducted from single channel audio.
Because utilizing a high number of speakers with PIT is difficult, our system was only trained to output the most dominant seven speakers even if the input contained more than seven speakers.
The results are shown in \autoref{tbl:dihard}.
Our SA-EEND with EDA achieved a DER of \SI{32.59}{\percent}, which outperformed the baseline \cite{sell2018diarization} and the best pre-is2019-deadline system by the DI-IT team \cite{novoselov2019speaker}, but it could not beat the best post-is2019-deadline system by the BUT team \cite{landini2020but}.
We note that our system is based on \SI{8}{\kHz} audio, while others use \SI{16}{\kHz} audio with additional training data from VoxCeleb datasets \cite{nagrani2020voxceleb}.
Evaluations on high-resolution audio with additional data are left for future work.

\section{Conclusions}
In this paper, we proposed EDA to calculate attractors from a sequence of embeddings, and we applied it to SA-EEND to implement end-to-end speaker diarization for speech mixtures of a flexible number of speakers.
Our method achieved state-of-the-art DERs on conditions including both a fixed and a flexible number of speakers.

\bibliographystyle{IEEEtran}
\bibliography{mybib}
\end{document}